# Topological kagome magnets and superconductors

Jia-Xin Yin[1,2]†, Biao Lian[1]†, M Zahid Hasan[1,3,4,5]†

[1]Department of Physics, Princeton University, Princeton, NJ, USA.

[2]Department of Physics, Southern University of Science and Technology, Shenzhen, Guangdong, China.

[3]Princeton Institute for the Science and Technology of Materials, Princeton University, Princeton, NJ, USA.

[4]Materials Sciences Division, Lawrence Berkeley National Laboratory, Berkeley, CA, USA.

[5]Quantum Science Center, Oak Ridge, TN, USA.

†Corresponding authors, E-mail:

yinjx@sustech.edu.cn; biao@princeton.edu; mzhasan@princeton.edu

**A kagome lattice naturally features Dirac fermions, flat bands and van Hove singularities in its electronic structure. The Dirac fermions encode topology, flat bands favour correlated phenomena such as magnetism, and van Hove singularities can lead to instabilities towards long-range many-body orders, altogether allowing for the realization and discovery of a series of topological kagome magnets and superconductors with exotic properties. Recent progress in exploring kagome materials has revealed rich emergent phenomena resulting from the quantum interactions between geometry, topology, spin and correlation. Here we review these key developments in this field, starting from the fundamental concepts of a kagome lattice, to the realizations of Chern and Weyl topological magnetism, to various flat-band many-body correlations, and then to the puzzles of unconventional charge-density waves and superconductivity. We highlight the connection between theoretical ideas and experimental observations, and the bond between quantum interactions within kagome magnets and kagome superconductors, as well as their relation to the concepts in topological insulators, topological superconductors, Weyl semimetals and high-temperature superconductors. These developments broadly bridge topological quantum physics and correlated many-body physics in a wide range of bulk materials and substantially advance the frontier of topological quantum matter.**

**Cradle of kagome electrons**

A kagome lattice (Fig, 1**a**), made of corner-sharing triangles, is a geometrically frustrated two-dimensional (2D) lattice introduced to quantum physics[1] in 1951, as an analogy to a type of bamboo baskets used in eastern countries. Interestingly, the kagome pattern has no direct analogy to wild nature, while a similar pattern has long been used as the star of David in religious ceremonies and Hexagram in alchemy symbols. In the quantum physics research, inspired by Onsager's exact solution on the Ising model for square lattice[2] in 1940s, researchers extended the study of magnetic phase transitions to triangular, honeycomb, and eventually kagome lattices. In the 1951 kagome work[1], Syôzi demonstrated that, in contrast to the ferromagnetic case, phase transition does not occur in the kagome lattice with the nearest neighbor antiferromagnetic Ising interaction. Nowadays, under antiferromagnetic exchange interaction, the geometrical spin frustration in kagome lattice has been widely appreciated, and it shows a great potential to realize the quantum spin liquids states[3] that feature long-range quantum entanglement, fractionalized excitations, and the absence of ordinary magnetic order[4]. While these studies are about Heisenberg spins in the kagome lattice, there has long been a debate about quantum magnetism since the 1930s: Heisenberg's view that magnetic moments come from electrons localized in atoms[5], or Stoner's view that magnetization collectively arises from itinerant electrons[6]. Under this broad context, in the 1991 kagome work[7], Mielke studied the Hubbard model of kagome lattice, and demonstrated that the flat band in its electronic structure stabilizes the ferromagnetic ground state. Finally, through those early quantum magnetism research, the electronic structure of the kagome lattice is unveiled and began to reveal its power of emergence[8].

**Fundamental concepts for kagome electrons**

For a kagome lattice electron tight-binding Hamiltonian $H = -\sum_{\langle ij \rangle, s_z = \pm\frac{1}{2}} (t c_{i,s_z}^\dagger c_{j,s_z} + h.c.)$ with a real nearest neighbor (denoted by $\langle ij \rangle$) hopping $-t$ ($c_{i,s_z}^\dagger$ and $c_{i,s_z}$ : spin $s_z$ electron creation/annihilation operators on site $i$), its electronic band structure exhibits an exactly flat band at energy $2t$ in the whole Brillouin zone, in quadratic band touching with the neighboring band (Fig. 1**b**). This flat band originates from sub-dimensional eigen-wavefunctions with exact cancellation of hoppings due to lattice geometry, which can be generalized into a class of tight-binding models called the line graphs[7,9]. Moreover, the band structure features Dirac cones at the Brillouin zone corners, and two van Hove singularities at the Brillouin zone boundaries (Fig. 1**b** and **c**). Theoretical understanding of the emergent physics of kagome electrons deepens along with the introduction of theories on topological quantum matter[10-12] and high-temperature superconductivity type correlated many-body physics[13] into kagome systems.

A pioneering idea[14] is that, by a hopping flux $\phi \neq 0$ and $\pi$ in each triangle plaquette (Fig. 1**a**), energy gaps can be opened at the Dirac points between the two dispersive bands and the quadratic band touching point between the flat band and its neighboring band, which leads to a Chern number $C = \pm1, 0, \mp1$ in the three bands (in energetic order), in resemblance to the Haldane model in the honeycomb lattice[15]. Each occupied band then contributes a quantized Hall conductance $\sigma_{xy} = Ce^2/h$ (e is the elemental charge, and h is the Planck's constant). The flux $\phi$ amounts having a complex hopping $-t - i\lambda' = -|t'|e^{i\phi/3}$ counterclockwise in each triangle ($-\lambda'$ is the imaginary part of the complex hopping $-t'$), which can arise from the spin Berry phase in a kagome magnet with a chiral spin texture[14]. With the conceptual development of quantum spin Hall effect and $Z_2$ topological number[16,17], the Kane-Mele type of spin-orbit coupling is also introduced to kagome lattice[18,19,20] (Fig. 1**d**), which takes the form $H_{soc} = -\lambda \sum_{\langle i \to j \rangle, s_z = \pm\frac{1}{2}} 2s_z (i c_{i,s_z}^\dagger c_{j,s_z} + h.c.)$, where $\langle i \to j \rangle$ is nearest neighbor pointing counterclockwise in a triangle, and $\lambda$ is the spin-orbit coupling strength. It induces opposite fluxes $\pm\phi$ for opposite spins $s_z = \pm\frac{1}{2}$ and opens spin Chern gaps. The gap opened at either the Dirac points or the flat band quadratic band touching is $Z_2$ topologically nontrivial (quantum spin Hall), which will develop helical edge states protected by the time-reversal symmetry. By further adding an out-of-plane magnetization $H_{M_z} = -M_z \sum_{i,s_z=\pm\frac{1}{2}} s_z c^\dagger_{i,s_z} c_{i,s_z}$ to the kagome lattice ($M_z$ is the z-direction Zeeman field) that lifts the spin $s = \pm\frac{1}{2}$ degeneracy, the $Z_2$ topological gap becomes a Chern gap, and the kagome lattice will carry a chiral edge state correspondingly. In kagome lattice materials, magnetism often arises spontaneously, making such a scenario possible. When the Fermi level is filled within the Chern gap, the kagome lattice exhibits a quantized anomalous Hall effect[20]. When the electron filling is away from the Chern gap, the total Berry curvature below the Fermi level will contribute to the anomalous Hall conductivity $\sigma_{xy} \approx \frac{\Delta}{2|E_D|} \frac{e^2}{h}$, where $\Delta$ is the Chern gap size (twice the Dirac mass) and $E_D$ is the Dirac Fermi energy measured from the mid-gap energy. When the sizes of $\Delta$ and $E_D$ are comparable, the Berry curvature effect is still substantial, leading to a giant or large anomalous Hall effect that is often reported for topological kagome magnets. Moreover, since the Kane-Mele type spin-orbit coupling $H_{soc}$ is proportional to the out-of-plane spin $s_z$, a sufficiently large in-plane magnetization $H_{M_x} = -M_x \sum_{i,s_z,s_{z'}} c^\dagger_{i,s_z} (\sigma_x)_{s_z,s_{z'}} c_{i,s_z}$ ($M_x$ is the x-direction Zeeman field) will polarize spins in-plane, making $H_{soc}$ effectively zero and closing the topological gap. This spin-orbit tunability leads to the magnetization direction control of the quantum state topology of the kagome lattice, which is broadly discussed in topological kagome materials with soft magnetism. Similar topological physics induced by spin-orbital coupling and magnetism can also occur when the Fermi level is between the flat band and its neighboring band. In reality, most kagome materials are 3D materials consisting of stacked kagome lattice layers, and the above physics of kagome lattice emerges in each layer in the weak interlayer coupling regime. Such layered kagome materials with the Fermi level in the Chern gap will exhibit the 3D quantum anomalous Hall effect[21]. When the interlayer coupling is strong, the dispersion can have a strong z-direction (perpendicular to layers) momentum $k_z$ dependence. If one views the band structure at each fixed $k_z$ as a 2D kagome system, its Chern gap can also undergo gap closings as a function of $k_z$, which correspond to Weyl points in the 3D Brillouin zone[22-24].

In parallel to the topological bands, the many-body interaction of kagome electrons is another research frontier for kagome materials. Since the discovery of copper-based and iron-based high-temperature superconductivity, it has been widely appreciated that many-body correlation lies at the heart of most exotic quantum phases of matter[13]. For the kagome electrons, the initial interest is in the realization of the flat band. Benefiting from the quenched kinetic energy, flat bands can host correlated electronic states, including ferromagnetism[7] that is often detected in kagome lattice materials. As discussed earlier, the spin-orbit coupling can open a Chern gap at both the Dirac cone and the quadratic band touching point of the flat band in the magnetic kagome lattice. At a partial filling of the flat band, the kagome lattice with interactions has the potential of realizing fractionalized topological phases, such as the fractional Chern insulator (fractional quantum Hall states without magnetic field)[19,25-27] (Fig. 1**e**). Soon after these flat band concepts are established, substantial interest is focused on the possible electronic orders at the van Hove filling[28-30], where the Fermi surface is nested and has saddle points on the edges of the Brillouin zone. The real-space superlattice for the nesting vector gives a 2×2 unit cell enlarging, and various kinds of spin density wave, charge density wave, and nematicity order with the similar wave vectors have been proposed[28-30]. These orders also serve as a precursor of unconventional superconductivity in the kagome lattice and are broadly relevant to the kagome superconductors. One intriguing charge order under the consideration of extended Coulomb interaction is one that leads to orbital currents and breaks time-reversal symmetry[30]. Similar orbital currents scenario has also been proposed for the kagome flat band full filling that leads to quantum anomalous Hall effect[31]. More broadly, the orbital currents physics in kagome lattice resembles the loop current proposal for cuprates as an attempt to understand the pseudogap physics[32], and effectively realizes the Haldane model phase[15]. Compared with the spin-orbit coupling driven topology, the interaction driven topology (defined as topology from order induced by many-body interaction of the active band electrons) may open the topological gap directly at the Fermi level (Fig. 1**f**). In addition, the time-reversal symmetry-breaking charge order links the kagome magnetism and kagome superconductivity. Actually, the initially proposed kagome superconductivity is through doping a kagome spin liquid[33] (Fig. 1**g**), where the superconducting order breaks the time-reversal symmetry.

In realizing these concepts, advanced techniques are applied in probing kagome electrons. While magneto-transport can probe the phase transition and Berry curvature, it is only sensitive to the electronic states at the Fermi level. While angle-resolved photoemission probes the band dispersions, it lacks magnetic tunability and integrates different magnetic domains. While scanning tunneling microscopy can probe the local density of states under a magnetic field, it is only sensitive to electronic states of the surface kagome layer. While scattering techniques can probe the bulk spin and lattice order, they often lack charge and orbital sensitivity. While first-principles calculation is very powerful in identifying the topological index of the electronic structure, correlations associated with magnetism lead to non-rigid band shift so that the chemical potential and band renormalization require corrections in reference to the experimental data. Therefore, these complementary techniques need to cooperate to establish the topological or many-body effects of kagome electrons.

**Chern quantum phase and spin-orbit tunability**

In the research of topological magnets, several transition metal based kagome materials come to researchers' attention, initially owing to their large anomalous Hall effect[34-37]. As learned from the quantum Hall effect, the topological nature of the anomalous Hall effect (i.e. intrinsic Berry curvature contribution) had readily been widely appreciated[38] at the time of these studies in kagome materials, especially after the experimental realization of the quantum anomalous Hall effect[39]. One natural explanation would be to associate the anomalous Hall effect with the Berry curvature near the topological gap in the spin-orbit coupled kagome bands. In ferromagnet $Fe_3Sn_2$, photoemission and tunneling experiments have detected Dirac-like gaps[40,41] and a kagome bilayer electronic structure is identified. However, $Fe_3Sn_2$ is a soft magnet with no zero-field anomalous Hall effect, and is known to exhibit an in-plane magnetization ground state with complex spin textures leading to skymions and topological Hall effects[42-44]. While this situation does not ideally match with a kagome Chern magnet picture, its soft magnetism and electron correlation nature leads to giant and anisotropic spin-orbit tunability of great interest that we discuss later[41,45-48]. Another concern is that most topological kagome magnets have an Sn atom in the center of their kagome lattice unit, which could lead to additional hopping that affects the ideal kagome band structure.

An alternative family[49-51] $RMn_6Sn_6$ (R = rare earth element) is then proposed to overcome these two problems. The large chemical pressure of rare earth can push the Sn atom away from the kagome layer. Particularly, ferrimagnet $TbMn_6Sn_6$ stands out due to its unique out-of-plane magnetization required for Chern magnetism (Fig. 2**a**). Scanning tunneling spectroscopy shows that the Mn kagome lattice exhibits distinct Landau quantization under a magnetic field, which features a spin-polarized Dirac dispersion with a Chern gap $\Delta = 34$ meV with Dirac Fermi energy $E_D = 130$ meV. This Landau quantization fan maps out the Chern gapped Dirac dispersion, which agrees well with angle-resolved photoemission results of the occupied states. Furthermore, a topological edge state with substantially reduced quasi-particle scattering is detected at 130meV, which is right within the Chern gap. Similar to the detection of spectroscopic Landau quantization, the longitudinal transport detects topological quantum oscillations[51,52] (Fig. 2**b** top panel) in $RMn_6Sn_6$. The quantum oscillation matches with the Dirac Fermi surface from spectroscopic probes[50], and shows a cyclotron mass agreeing with the spectroscopic data. Intrinsic anomalous Hall conductivity (Fig. 2**b** bottom panel) per kagome layer is measured to be $\sigma_{xy} = 0.14 e^2/h$, agreeing with the estimation of Berry curvature contribution based on spectroscopic data $\sigma_{xy} = \Delta/(2E_D) \times e^2/h = 0.13e^2/h$. These quantitative spectroscopic-transport agreements build up the bulk-boundary-Berry correspondence for the kagome Chern magnet (Fig. 2**a**). Owing to the strong out-of-plane magnetism and large Berry curvature from Dirac fermions with a Chern gap, $TbMn_6Sn_6$ exhibits zero-field anomalous Hall, anomalous Nernst, and anomalous thermal Hall effect[51-53] (Fig. 2**b** bottom panel). The charge-entropy scaling derived from these anomalous transverse transport goes beyond the Mott formula[54] and the Wiedemann-Franz law[55] for ordinary electrons, but can be described, interestingly, by the new scaling behavior derived from Dirac fermions with a Chern gap suggesting a topological behavior consistent with spectroscopic observations[50,52].

Owing to the pristine kagome lattice structure, kagome type band structure and anomalous Hall effect are widely searched and detected in $RMn_6Sn_6$ and closely related compounds[50,56-64]. The newly discovered $RV_6Sn_6$ presents a complementary case where V-based kagome layer is paramagnetic[60] and R layer can be magnetic, and their clean kagome type band structure is well captured by photoemission experiment and explained by the first-principles. The diverse chemical compositions and magnetic structures of these materials naturally lead to large spin-orbit tunability of its topological electronic structure. For instance, the rare earth element can act as a tuning knob of the Chern gap, which scales with the underlying de Gennes factor[51]. In another related ferromagnet $LiMn_6Sn_6$, a larger anomalous Hall effect is identified, which is likely due to a smaller Dirac Fermi energy[63]. As discussed in the fundamental concept of kagome electrons, in-plane magnetization tends to close the Chern gap, leading to giant spin-orbit tunability with respect to different magnetization configurations. Such magnetization control has been discussed in $Fe_3Sn_2$ system[41,45-48] and $RMn_6Sn_6$ systems[60,63,64] via spectroscopic and transport experiments. Another intriguing phenomenon is the detection of nematicity in $Fe_3Sn_2$[41,45]. When rotating its magnetization along different in-plane directions, Zeeman energy shift and quasi-particle scattering signal both exhibit nematicity (two-fold symmetry, see Fig. 2**c** and **d**), a hallmark of correlated electrons. The tunability of the kagome Chern magnets reveals a strong interplay between the exchange field or external magnetic field, electronic excitations, and nematicity, providing new ways of controlling spin-orbit properties.

**Weyl fermion and antiferromagnetic spintronics**

Weyl fermions were first introduced in 1929 in the context of high-energy physics as massless particles with definite chirality, described by a two-component spinor obeying the Weyl equation[22]. Weyl fermions arise as low-energy quasi-particle excitations in 3D crystals lacking inversion symmetry or time-reversal symmetry[23,24,65,66]. In a Weyl crystal, Weyl fermions are the sources and sinks of the Berry curvature in momentum space (Fig. 3**a**). On its surface, a topological Fermi arc surface state is formed, which connects the projections of two Weyl points in the momentum space of the surface. The emergence of Weyl fermions in kagome magnets naturally requires strong interlayer coupling, such that the bulk electron bands are 3D. Noncollinear antiferromagnet $Mn_3X$ (X = Sn, Ge) and ferromagnet $Co_3Sn_2S_2$ stand out[35-37,67-69], as the direct kagome lattice stacking in $Mn_3X$ and strong ionic bonding through $S^{2-}$ in $Co_3Sn_2S_2$ enhance the interlayer coupling. Based on the noncollinear in-plane antiferromagnetism[70] in $Mn_3Sn$ and out-of-plane ferromagnetism[71] in $Co_3Sn_2S_2$, their Weyl fermions have been identified through theoretical calculation[35-37,68,69] (Fig. 3**b**). In these materials, unique negative magnetoresistance (Fig. 3**c**) has been reported, which is consistent with the expected chiral anomaly for Weyl fermions[36,69]. Meanwhile, certain features of the 2D kagome electrons are still tracible in their band structure, such as their kagome flat bands[72,73].

One most remarkable property of $Mn_3Sn$ is its room-temperature zero-field anomalous Hall effect[35] (and other related Berry curvature effects[74-80]) under a weak and soft magnetic moment, providing exceptional application value for antiferromagnetic spintronics[81]. For instance, in a bilayer device made with nonmagnetic metals and the $Mn_3Sn$ film, an electric current can manipulate the Hall voltage by switching the chiral kagome antiferromagnetism[78] (Fig. 3**d**). The electronic band dispersions of $Mn_3X$, however, have not been clearly detected, possibility owing to a strong electron correlation[69,73]. While similarly exhibiting giant anomalous Hall and related effects[36,37,82-88], $Co_3Sn_2S_2$ features a large ratio between anomalous Hall conductivity and longitudinal conductivity[36]. Such improvement is likely due to its cleaner band structure that also enables better spectroscopic characterizations of its Weyl topology[82]. Guided by the first-principles calculation, progress is progressively made through counting the Chern numbers of the surface band crossings[83], imaging the linear band crossing with surface doping[84], imaging the diversity of quasi-particle interferences on surfaces hosting Fermi arcs[85], imaging the magnetic phase transition of the topological band structure[87], and the detection of the spin-orbit coupling induced energy gap[86,88].

We further clarify the relationship between anomalous Hall effect and Weyl fermions. A pioneering study[89] of anomalous Hall in $Mn_3X$ type of materials attributes it to the Berry curvature effects from noncollinear magnetism without inviting the Weyl fermions. In magnetic crystals, there are generic magnetic nodal lines in the bulk bands protected by spinless mirror symmetry (Fig. 3e), and the spin-orbit coupling opens anisotropic gaps along the nodal lines, yielding nodes as pairs of Weyl cones. As one pair of Weyl fermions are often at the same energy acting as source and sink of Berry curvatures, respectively, their local integrated contribution to anomalous Hall conductivity essentially cancels out. On the other hand, the portions of the magnetic nodal lines that carry a large spin-orbit gap contribute to the integrated Berry curvature that leads to the anomalous Hall effect. This can be thought as a long-distance dipole effect of Weyl fermions[90], or as an effect of a local effective Chern gap. One initial evaluation of the spin-orbit gap in kagome Weyl magnets is to investigate the local geometrical impurity induced resonances[86] (Fig. 3f). For triplet geometrical impurity induced resonances, the spin-orbit coupling similarly breaks mirror symmetry to introduce a splitting gap[86], which is of the similar size as the magnet nodal line gap[88].

**Flat band correlation and many-body physics**

Kagome lattices host both high-velocity electrons (Dirac or Weyl fermions) and vanishing velocity electrons (flat band). Flat band physics can be traced back to the fractional quantum Hall effect[25-27]. The Landau levels generated by a large magnetic field are perfect flat bands, and a fractional filling can produce the fractional quantum Hall effect with anyonic excitations. Without the magnetic field, flat bands are rare and emerge only in a few systems such as heavy-fermion compounds[91], kagome lattices (or line-graphs[7,9]) and twisted bilayer graphene[92,93]. Paramagnet CoSn stands out owing to its simple crystal structure and cleaner flat bands in spectroscopic experiments matching first-principles calculations[94-100]. The hallmark of kagome flat band — quadratic band touching in momentum space (Fig. 4a) — is visualized by angle-resolved photoemission experiments[94,96]. As originally pointed out in the modeling of kagome electrons[20], the highly anisotropic $d$ orbital can form flat bands. The $d$ orbitals uniquely arrange themselves in the hexagonal ring of the kagome lattice (Fig. 4a inset), leading to a nearly perfect canceling of their propagating wave functions[96]. A similar arrangement of lattice vibration is found to produce localized phonon modes — flat band phonons[95] (Fig. 4b).

The existence of flat band leads to several emergent effects. The kagome flat bands have been known to favor ferromagnetism[1,7]. The flat bands in ferromagnet $Fe_3Sn_2$ and antiferromagnet FeSn can be related to their strong in-plane ferromagnetism[101-103]. With spin-orbit coupling, the quadratic degenerate points open a $Z_2$ topological gap (Fig. 4c), which is visualized in CoSn[94,96]. With the further introduction of the out-of-plane ferromagnetism, it turns into a Chern gap, as in the flat band case[72] of $Co_3Sn_2S_2$. The associated Berry curvature of this Chern gap will lead to sizable orbital magnetism[104,105]. Tunneling spectroscopy of the flat band peak shifts anomalously with both positive and negative magnetic field[72], which goes beyond the conventional Zeeman effect. The anomalous Zeeman shift further supports the existence of orbital magnetism, with its direction being opposite to the bulk magnetization direction (Fig. 4c), which is consistent with the first-principles calculation[72]. Similar negative orbital magnetism physics is also detected for impurity resonances[86,106] as local flat bands in $Co_3Sn_2S_2$ and for anisotropic magnetic susceptibility[99] in CoSn. The negative orbital magnetism associated with Dirac fermions is also detected in a momentum-resolved way[61] in antiferromagnet $YMn_6Sn_6$. Moreover, different from the bulk kagome flat band featuring a quadratic band touching, the surface flat band realized in FeSn does not have such a feature[103,107], while the realization of atomic film of FeSn makes it an appealing platform for flat band based spintronic devices[108,109].

Besides the spin-orbit driven topology, flat band can lead to strong many-body interactions. The itinerant electrons in CoSn sense the strong coupling to its phonon flat band[95], leading to a kink in the band dispersion (Fig. 4**d**) and peak-dip-hump structure in the tunneling spectrum. The electron-phonon coupling can be a factor that leads to competing electronic instabilities that result in its paramagnetic state instead of ferromagnetism. In the kagome magnet $Mn_3Sn$, its electron correlation is substantially strong, as indicated by the large band renormalization factor[69]. Under this strong correlation, its single-particle flat band can couple to the itinerant electrons at Fermi level, leading to a many-body resonance at the Fermi level[73] (Fig. 4**e**). The spectroscopic features of this many-body resonance, including Fano line shape and fast temperature broadening, resemble those detected in Kondo systems[91].

**Unconventional charge density waves**

The interplay between topology and interaction is extensively discussed in kagome superconductors. While kagome superconductors with a transition temperature $T_C$ up to 7K have been discovered for over half a century (such as $CeRu_2$ and $LaRu_3Si_2$)[110-113], recently discovered kagome superconductors[114-117] $AV_3Sb_5$ (A = K, Rb, Cs) exhibit a charge density wave[115,118,119], the ordering temperature T* of which reaches to 100K. Above T*, the band structure of $AV_3Sb_5$ features van Hove singularities at the Fermi level, and an indirect topological band gap[115] under consideration of spin-orbit coupling. Both magnetization and transport measurements find an anomaly at T*, which is considered as a possible signature of orbital ordering[114] or a charge density wave like instability[115]. Scanning tunneling microscopy identifies the 2 × 2 in-plane supercells of the charge modulation, charge energy gap, and charge density reversal across the gap below T*, which reveal a charge density wave order[118]. X-ray scattering further identifies the 2 × 2× 2 3D supercells[119] of below T*. These initial experiments establish the bulk 2 × 2 × 2 charge density wave as the many-body state in topological kagome superconductor $AV_3Sb_5$ below T* (Fig. 5**a**), and its in-plane wave vector agrees with the early studies of the possible electronic instability for kagome lattices[29,30]. The charge density wave with a similar 2 × 2 × 2 supercells has been detected in transition metal dichalcogenides[120], where chirality and nematicity of the order are discussed[120-123]. The antiphase coupling of the adjacent 2 × 2 order leads to nematicity in the first order and can feature chirality as a further orbital ordering[120-123]. Similar additional spatial symmetry breaking, as well as the associated lattice[124-134] and orbital[135-149] manifestations below the charge density wave state of $AV_3Sb_5$, is intensively investigated through various techniques[150]. Many of them connect with each other reasonably. For instance, tunneling[118,136] and photoemission[142,146,147] show consistent charge density wave gap for occupied states as of ~20meV (Fig. 5**b**). This systematic progress set the fundamentals to understand the dual lattice-orbital nature of the charge density wave in the kagome lattice.

An alternative paradigm in looking at the charge order focuses on its time-reversal symmetry breaking (Fig. 5**c**), which goes beyond conventional charge density wave but features orbital current physics as initially proposed for achieving the quantum anomalous Hall effect[15] and for the hidden phase in cuprates[32]. Pioneering magnetization[114], muon spin rotation[151] and tunneling[118] studies show that both the phase transition temperature and energy gap are insensitive to the magnetic field, and that there is no spin ordering. Thus, any magnetism should be a higher-order effect. One breakthrough is the detection of magnetic field switching of the 2 × 2 charge order chirality[118,135,136], while regions with defects or strains featuring a much smaller energy gap do not show chirality[118,134-136]. The phenomenon of magnetic field controlled chirality can be modeled[118] by considering a relative phase between three sets of order parameters describing the charge order in the three kagome sublattices. This complex order parameter breaks time-reversal symmetry, leads to a large anomalous Hall effect[152,153], and introduces a small orbital magnetization. Theoretical studies[154-156] have uncovered that the time-reversal symmetry-breaking charge order is energetically favorable in the kagome lattice close to the van Hove filling. This exotic charge order provides a rare example of many-body interaction-driven topology (Fig. 1**f**).

Magnetic moment sensitive probes[157-160], including muon spin rotation and Kerr rotation, further confirm the magnetic feature of the charge order. An external magnetic field can substantially boost the time-reversal symmetry breaking muon signal[157-159], providing insights beyond the initially considered magnetic field insensitivity[114]. Recent observation of field-tuned chiral transport[161] further elaborates how spatial chirality and time-reversal symmetry-breaking twist with each other to introduce nonlinear magneto-transport, which is consistent with the spectroscopy detection of field-tuned electronic chirality[118,135,136]. In connection to the kagome superconductor, charge order is detected[162,163] in kagome antiferromagnet FeGe. The essential physical picture of the kagome charge order is further tested, including van Hove singularity driven $2 \times 2$ charge instability, the anomalous contribution to the Hall effect, magnetic coupling of the charge order, and internal exchange magnetic field tuned chirality switching. In addition, robust edge states are detected within the charge order energy gap, agreeing with the topological nature of the time-reversal symmetry-breaking charge order (Fig. 1**f**).

**Magnetically intertwined superconductivity**

Superconductivity in kagome lattices has long been identified to coexist with magnetic phases. One earliest known superconductor hosting kagome lattice, $CeRu_2$, has been termed as a ferromagnetic superconductor[110] by Matthias. The normal state of kagome superconductor $CeRu_2$ is a weak magnet, likely owing to the flat band correlation[110,111]. Coincidently, one early theory for kagome superconductors (Fig. 1**g**) predicts a time-reversal symmetry-breaking superconducting ground state[33]. The intertwining between magnetism and superconductivity has also been theoretically considered for kagome electrons at the van Hove filling[28-30] and the Dirac filling[164]. These pioneering works set up a tone for exploring magnetically intertwined kagome superconductivity.

Putting this research theme in $AV_3Sb_5$, both physical pressure and chemical doping can suppress the charge order, while superconductivity immediately gets enhanced[165-172], signaling their competition (Fig. 5**d**). A question is whether the time-reversal symmetry is broken for superconductivity. Initial muon study argues that this symmetry-breaking persists into superconductivity[157], since there is no additional relaxation rate change across $T_C$. This proposal is further substantiated by the observation of the relaxation rate change across $T_C$ through killing the charge order by pressure tuning[158]. Another evidence for such intertwined physics is to examine the effect of interactions impacting both the time-reversal symmetry breaking features of charge order and superconductivity. Theoretical analysis[30,154-156] indicate that the extended Coulomb interaction drives time-reversal symmetry-breaking charge order. Owing to the poorly screened Coulomb interaction, the charge number fluctuations are suppressed. According to the number-phase uncertainty, quantum phase fluctuations will proliferate leading to smaller superfluid density[173]. The penetration depth $\lambda$ of several kagome superconductors has been measured[111,113,157-159], where $\lambda^{-2}$ is proportional to the superfluid density, and suggests their small superfluid density. The scaling analysis of $T_C$ versus $\lambda^{-2}$ in these studies shows a markable difference from conventional superconductors but a close analogy to cuprates (Fig. 5**e**). The large $T_C/\lambda^{-2}$ value attests to the unconventional nature of kagome superconductivity in the correlation sense, and implies a connection with the underlying interactions driving time-reversal symmetry-breaking charge order.

In parallel, substantial research efforts[28-30,111,113,124,126,136,138,157,158,174-180] have also been devoted to determining the electron Cooper pairing symmetry[150]. Experimentally, both nodeless and nodal superconducting gap structure, as well as their quantum tuning, have been reported. However, a direct momentum space determination of the pairing symmetry and topological order have been lacking. Topological nature of the superconducting state remains one of the active areas of investigation. Experiments have shown emergent features of the electron pairing, including the observation of putative Majorana zero mode[124] and detection of pair density waves[126] on the kagome superconductor surface.

**Future opportunities**

Accompanied by the rise of interest in quantum material studies[181], quantum interactions in topological kagome magnets and related superconductors effectively serve as pairing glue for communities of correlated electrons and topological matter. This review aims not to conclude this field but to invite more to explore and enjoy this quantum party held in kagome lattice. At times, history mirrors the future. Imagine that we were in 2012 when substantial conceptual literature for kagome physics had existed, then could we predict these exciting discoveries reviewed here? The answer is unlikely to be positive, but it reflects the fast-growing nature of this field: it appears to be full of opportunities and seem to be progressing beyond projective expectations.

For the kagome Chern magnet such as $TbMn_6Sn_6$ and similar materials the corresponding mono-layer or few-layers device has not been successfully fabricated, which would allow for the demonstration of their cleaner electronic band structures and quantized transports. First-principles guided materials search towards more isolated or distinct kagome bands[20] is still needed for further development. One alternative route is to design and grow surface-supported metal-organic kagome frameworks[182,183]. For the kagome Weyl magnet $Mn_3Sn$, the experimental visualization of its Fermi arc and its vector magnetization control remain unsettled. Similarly, several kagome magnets exhibiting large transverse anomalous transport[184,185], including $UCo_{0.8}Ru_{0.2}Al$ and $Fe_3Sn$, still await spectroscopic visualization of their topological fermions. From the first-principles prediction[186] and microscopic studies on bulk crystal's surface edge[187], the monolayer of $Co_3Sn_2S_2$ is expected to exhibit the Chern insulator phase, which awaits experimental realization. The realization of monolayer or few-layer kagome magnets would also be a significant step toward the prediction of the high-temperature fractional Chern insulator phase[19], which requires an isolated flat Chern band with strong interaction and electrically tunability. A new opportunity has been provided by computational advances in building a catalogue of flat-band stoichiometric materials[188]. Moreover, the Landau level of spin-orbit coupled kagome flat band has recently been shown to be highly unconventional[189]. The high-energy-resolution and high-magnetic-field based spectroscopic mapping of the Landau fan of kagome flat band in $Co_3Sn_2S_2$ and CoSn can serve as promising tools for testing this theoretical modeling. The identification[162,163] of a charge density wave in FeGe is encouraging for the search of charge order in more topological kagome magnets. In parallel, since the charge order can be a precursor of superconductivity, it is also meaningful to further engineer FeGe with chemical doping and pressure to look for superconductivity. In kagome superconductor $AV_3Sb_5$, electronic evidence for time-reversal symmetry-breaking superconducting state has been lacking. The search for time-reversal breaking superconductivity with odd parity pairing may allow us to realize intrinsic topological superconductivity in either 3D or mono/few-layer kagome materials with a sizeable gap, and the bulk Majorana zero modes and chiral Majorana edge states thus open a route to exploring the quantum information frontier[190,191].

While the above questions are of current interest, there are longer-term explorative directions. For instance, it has long been proposed that kagome lattice systems are promising for the realization of gapped or Dirac spin liquid[192,193]. The existence of time-reversal symmetry breaking in realistic materials may also facilitate the possibility of chiral spin liquid[194,195] with detectable quantized transport[196,197]. The doped kagome quantum spin liquid can become an exotic superconductor[33], and may give rise to more unknown phases in mixing the quantum spin liquid and the superconducting state. In addition, dopped kagome herbertsmithite systems can host strongly correlated Dirac fermions[164], and are recently proposed to realize viscous electron fluids that can be described by holography to make a possible model for quantum connection to gravity[198]. Moreover, kagome physics may occur in Moiré systems that are highly tunable. For example, the twisted kagome lattice bilayer can yield emergent flat bands[199] similar to that in the magic angle twisted bilayer graphene. On the contrary, twisted graphene system is also predicted to be able to produce kagome electronic bands[200], which may allow the realization of pristine kagome electrons without spin-orbital coupling and related interaction driven phases. While the existing research in quantum materials remains largely within the interplay of quantum (anomalous) Hall and (unconventional) superconductivity, we look forward to reaching a regime of "unknown unknown" with the next generation of cleaner and artificial kagome materials.

**Figures**

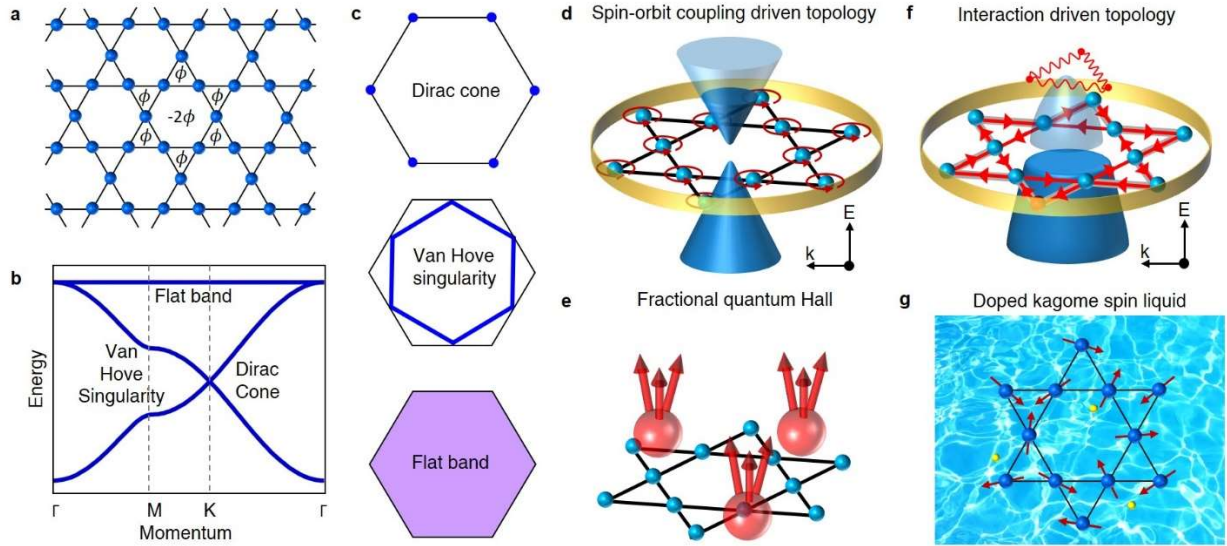

**Figure 1 Fundamental concepts for kagome electrons. a,** A kagome lattice made of corner-sharing triangles. Each triangle can carry a quantum flux ϕ, and each hexagon carries a quantum flux -2ϕ. **b,** Band structure for a kagome lattice considering the nearest neighbor electron hopping term with ϕ = 0. This band structure contains three key features: Dirac cones at K (or K') points, van Hove singularities at M points and a flat band in the momentum space. **c,** Contour of the Fermi surfaces in the momentum space at Dirac cone (upper panel), van Hove singularity (middle panel), and flat band (lower panel) energies, respectively. **d,** Atomic spin-orbit coupling (red circular arrows) driven topology in kagome lattice (blue spheres). The spin-orbit coupling opens a topological gap at the Dirac cone (blue cones) in the band structure (by producing a nonzero ϕ), and induces a topological edge state (yellow ring). The light (dark) blue cone represents unoccupied (occupied) electronic states, illustrating the generic case of Fermi energy away from the Dirac gap. **e,** Illustration of flat band driven fractional Chern insulator in the kagome lattice (blue spheres). The red spheres illustrate electrons bound with an odd number of Berry flux quanta in the ground state. **f,** Many-body interaction (red wavy lines) driven topology in the kagome lattice (blue spheres). The interaction (that can be related with the van Hove singularities) not only opens a charge gap at Fermi energy, but also produces a nontrivial Berry phase that leads to orbital currents (red arrows) and a topological edge state (yellow ring). The light and dark blue cones represent unoccupied and occupied electronic states, respectively. Compared with the spin-orbit coupling driven topology that often opens the topological gap away from Fermi energy, the interaction driven topology may open the topological gap directly at the Fermi level through long-range orders even without Dirac fermions. **g,** Doped kagome spin liquid as initially proposed for unconventional superconductivity in the kagome lattice. The blue spheres denote kagome lattice, the red arrows denote local spins, and the yellow spheres denote doped charge carriers.

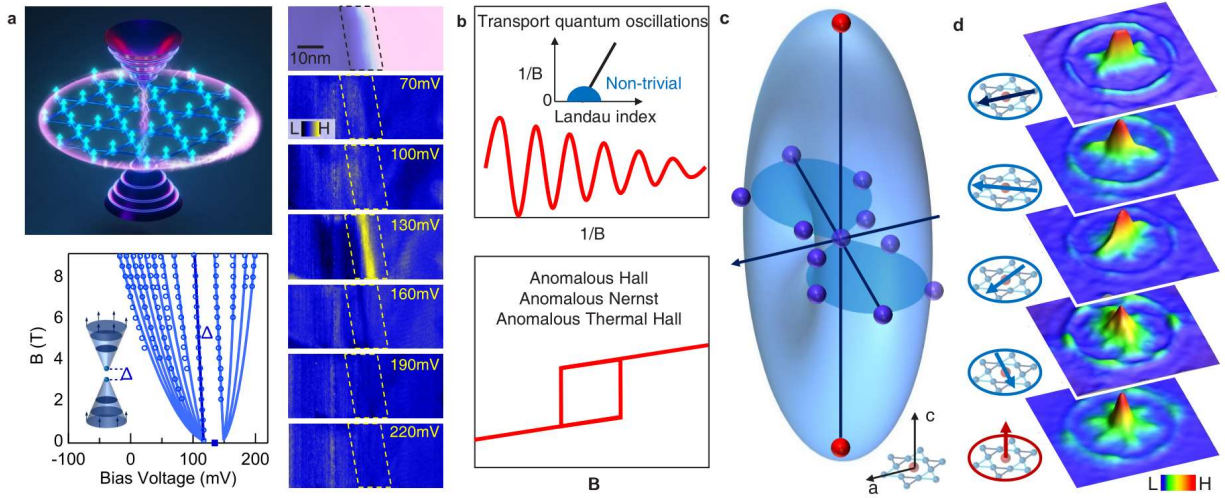

**Figure 2 Chern quantum phase and spin-orbit tunability. a,** Left top image illustrates the features of a kagome Chern magnet TbMn$_6$Sn$_6$ in real space and momentum space (in each 2D kagome layer). In momentum space, spin-polarized Dirac fermions with a Chern energy gap (two separated cones) can exhibit Landau quantization. In real space, the spin–orbit-coupled magnetic kagome lattice (spheres with arrows pointing up) carries a topological edge state within the Chern energy gap. Left down panel shows the fitting of the Landau fan data (circles) with the spin-polarized and Chern gapped Dirac dispersion (solid lines). Inset: schematic of Landau quantization of Chern gapped Dirac fermions. The right panel shows dI/dV maps taken at different energies across a step edge (top). The map taken within the Chern gap energy (130 meV) shows a pronounced topological edge state. **b,** Topological transport signal for kagome Chern magnet TbMn$_6$Sn$_6$. The upper panel illustrates the longitudinal transport quantum oscillations. The inset shows that the nontrivial Berry phase can be extracted from the intersection of the Landau index and 1/B=0, where the Landau index is extracted from the oscillation peaks. The lower panel illustrates the anomalous transverse transport. **c,** Vector-magnetization-induced energy shift of a quantum state in topological kagome magnet Fe$_3$Sn$_2$. The light-blue surface shows a 3D illustration of the nematic energy shift ΔE as a function of the magnetization vector, which exhibits a nodal line along the a-axis. **d,** Vector magnetization control of the electronic scattering in a topological kagome magnet Fe$_3$Sn$_2$. Electronic structure scattering patterns as a function of the magnetization direction, which is indicated in the insets with respect to the kagome lattice. The topmost QPI pattern shows the spontaneous nematicity along the a-axis. Magnetization along other directions can alter, and, thus, control, the electronic scattering symmetry. Panels **a** adapted from ref.50, Springer Nature Limited. Panels **c** and **d** adapted from ref.41, Springer Nature Limited.

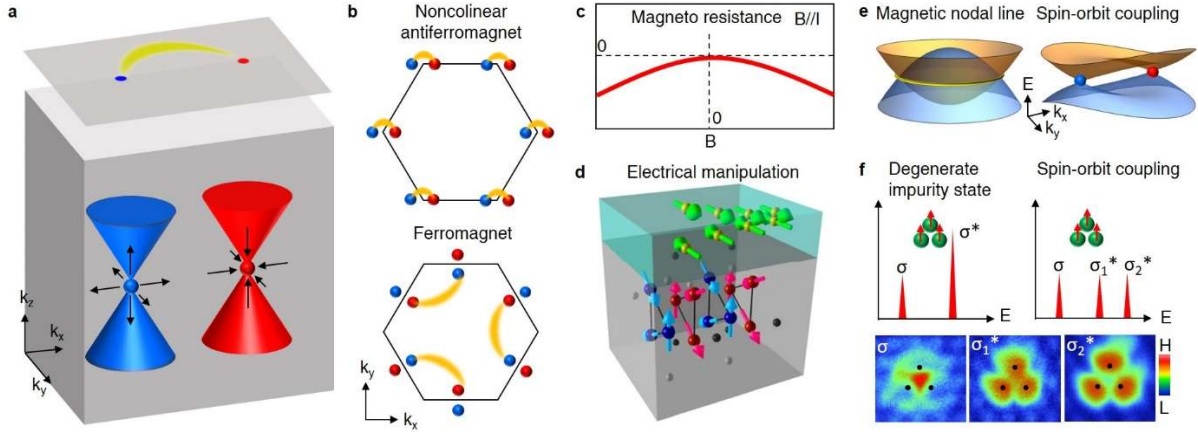

**Figure 3 Weyl fermion and antiferromagnetic spintronics. a,** Illustration of a pair of Weyl fermion quasi-particles with opposite chiral charges (red and blue spheres) in the 3D momentum space. The arrows illustrate the direction of the Berry curvature field in the vicinity of each Weyl fermion. The red and blue cones illustrate their respective linear Weyl dispersions in the bulk momentum space. On the top surface, a Fermi arc state (yellow arc) connects two Weyl cones in the surface momentum space. **b,** Illustration of Weyl Fermion positions and Fermi arc connectivity in the surface momentum space of a kagome noncolinear antiferromagnet $Mn_3Sn$ (up panel) and kagome ferromagnet $Co_3Sn_2S_2$ (lower panel). Owing to the rotation symmetry breaking of the noncolinear antiferromagnetism and spin-orbit coupling, the Weyl Fermion distribution and Fermi arc connectivity also break rotational symmetry. **c,** Negative magnetoresistance detected in $Mn_3Sn$ and $Co_3Sn_2S_2$ that is consistent with chiral anomaly expected for Weyl fermion quasi-particles. **d,** Electrical manipulation of the anomalous Hall effect for $Mn_3Sn$ films (lower layer) through injecting current into nonmagnetic metals (upper layer). **e,** Illustration of magnetic nodal line (yellow rings in the left image), and the spin-orbital coupling induced anisotropic splitting of the magnetic nodal ring (right image) that produces Weyl fermion quasi-particles (Weyl points at red and bule spheres). The blue and green shades illustrate the linear dispersion of the magnetic nodal line or Weyl cones in energy-momentum space. **f,** Real-space probe of the spin-orbit coupling induced quantum state splitting in $Co_3Sn_2S_2$. Three interacting magnetic impurity states with $C_{3V}$ geometry produce one bonding state σ and one doubly degenerate antibonding state σ* (top left panel). The spin-orbit coupling splits σ* into two states as $σ_1$* and $σ_2$* (top right panel), in analogy to the splitting of magnetic nodal line. The low panels show the real-space imaging of these three impurity-induced spin-orbit quantum states in $Co_3Sn_2S_2$. Panel **d** adapted from ref.78, Springer Nature Limited. Panel **f** adapted from ref.86, Springer Nature Limited.

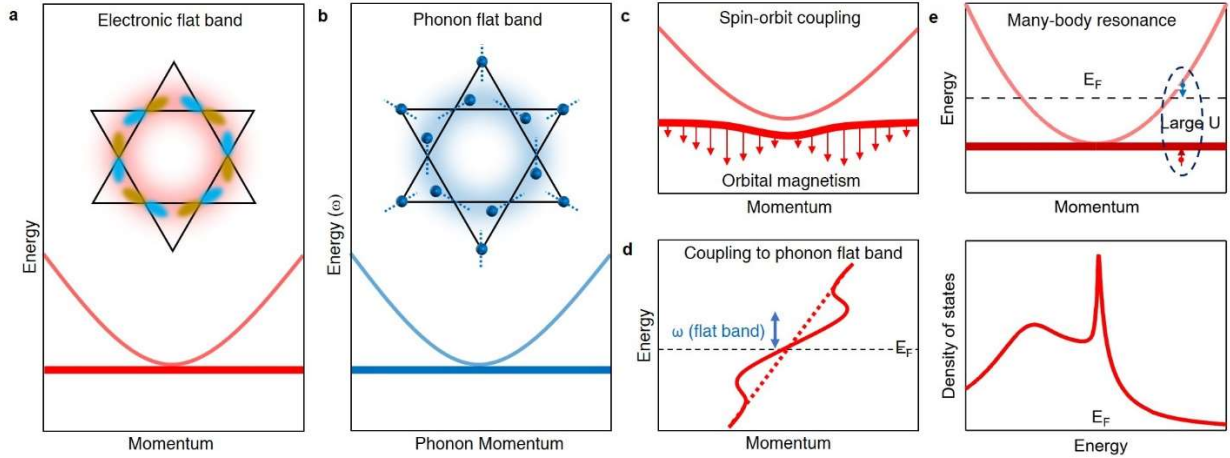

**Figure 4 Flat band correlation and many-body resonance. a,** The main panel displays the electronic flat band dispersion and how it touches a parabolic band without spin-orbit coupling, as discussed in kagome magnet $Co_3Sn_2S_2$ and kagome paramagnet CoSn. The inset shows the orbital configuration corresponding to the flat band in real-space (orbitals consider here are $d_{xz}$, $d_{yz}$ orbitals from top view). **b,** The main panel displays the phonon flat band and how it touches a parabolic band, as discussed in CoSn. The inset shows the lattice vibration distortion pattern corresponding to the phonon flat band (dashed lines indicate the vibration directions of each atom). **c,** Spin-orbit coupling effects, which include a gap opening and negative orbital magnetism of the flat band that are detected in $Co_3Sn_2S_2$ and CoSn. **d,** Electronic structure coupling to the flat band phonon, as detected in CoSn. The electron band exhibits a double kink feature at the energy of the phonon flat band. The electron-phonon coupling can lead to charge order and superconductivity instabilities. **e,** Many-body resonance owing to coupling of the flat band to the itinerant band under strong electron correlation (large U, as illustrated in the upper panel) as detected in kagome antiferromagnet $Mn_3Sn$. The lower panel illustrates the Fano-shaped resonance in the density of states at Fermi energy, $E_F$.

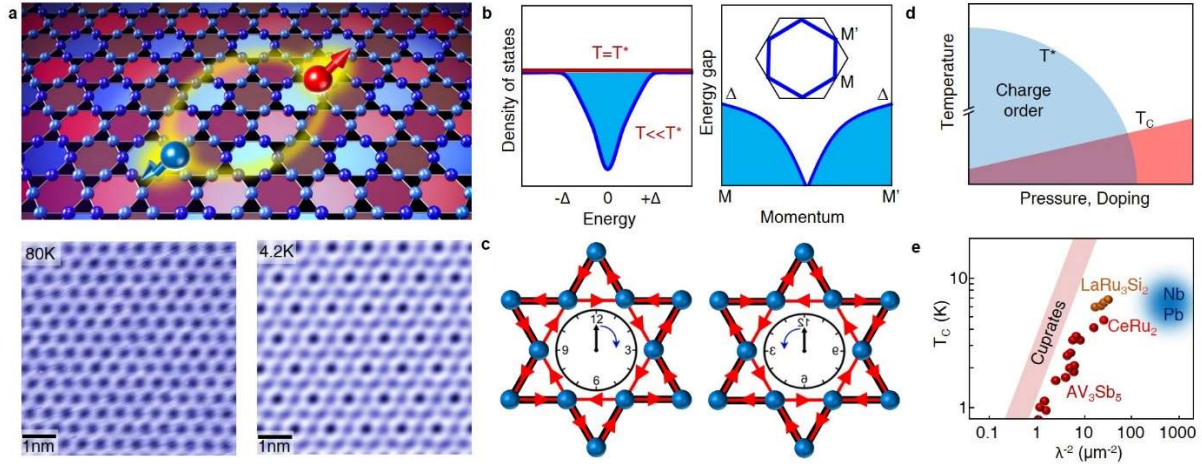

**Figure 5 Charge density wave and superconductivity. a,** The top panel is the schematic of the charge order and superconductivity detected in kagome superconductor $AV_3Sb_5$. The dark and light blue spheres form the kagome lattice. The shade of the color represents the unusual distribution pattern of the charge order. The large red and bule spheres with arrows represent Cooper pairing of the superconductivity. The lower panels show an example of charge order imaging. The topographic images are taken above (left) and below (right) the charge ordering temperature T* for kagome superconductor $KV_3Sb_5$, where a 2×2 supercell of the charge order is identified. **b,** Charge order gap in the density of states (left) and momentum electronic structure (right). A V-shaped charge order gap opens at the Fermi level below T* (left image); in momentum space the gap opens in an anisotropic way, with its maximum value at M or M' (right image, and its inset shows the Fermi surface). **c,** Time-reversal symmetry-breaking of charge order in $AV_3Sb_5$ and kagome antiferromagnet FeGe. The thick black lines mark the unit cell of the kagome charge order. The red arrows illustrate the orbital currents in the kagome lattice that breaks time-reversal symmetry. **d,** Physical pressure or chemical doping based phase diagram, showing competition between charge order and superconductivity. **e,** Scaling plot for superconducting transition temperature $T_C$ versus inverse squared magnetic penetration depth. The scaling relation of kagome superconductors (including $LaRu_3Si_2$, $AV_3Sb_5$ and $CeRu_2$) is close to that of cuprates but away from that of conventional superconductors such as Nb and Pb. Panel **a** adapted from ref.118, Springer Nature Limited.

**Acknowledgements**

We thank our research collaborators for various discussions on kagome physics. M.Z.H. acknowledges support from the US Department of Energy, Office of Science, National Quantum Information Science Research Centers, Quantum Science Center and Princeton University; visiting scientist support at Berkeley Lab (Lawrence Berkeley National Laboratory) during the early phases of this work; support from the Gordon and Betty Moore Foundation (GBMF9461) for the STM and the theory work; and support from the US DOE under the Basic Energy Sciences programme (grant number DOE/BES DE-FG-02-05ER46200) for the theory and angle-resolved photoemission spectroscopy work. B.L. is supported by the Alfred P. Sloan Foundation, the National Science Foundation through Princeton University's Materials Research Science and Engineering Center DMR-2011750; and the National Science Foundation under award DMR-2141966. J.-X.Y. acknowledges support from Princeton University, as well as the support from South University of Science and Technology of China principal research grant (number Y01202500). M.Z.H. also acknowledges visiting scientist support from Stanford University during the last phase of this work.

**Contributions**

All authors discussed the content of the manuscript, and reviewed and edited the entire manuscript.